# Iterative Algorithms for Joint Scatter and Attenuation Estimation From Broken Ray Transform Data

Michael R. Walker II*, *Member, IEEE*, and Joseph A. O'Sullivan, *Fellow, IEEE*

*Abstract*—The single-scatter approximation is fundamental in many tomographic imaging problems including x-ray scatter imaging and optical scatter imaging for certain media. In all cases, noisy measurements are affected by both local scatter events and nonlocal attenuation. Prior works focus on reconstructing one of two images: scatter density or total attenuation. However, both images are media specific and useful for object identification.

Nonlocal effects of the attenuation image on the data are summarized by the broken ray transform (BRT). While analytic inversion formulas exist, poor conditioning of the inverse problem is only exacerbated by noisy measurements and sampling errors. This has motivated interest in the related star transforms incorporating BRT measurements from multiple source-detector pairs. However, all analytic methods operate on the log of the data. For media comprising regions with no scatter a new approach is required.

We are the first to present a joint estimation algorithm based on Poisson data models for a single-scatter measurement geometry. Monotonic reduction of the log-likelihood function is guaranteed for our iterative algorithm while alternating image updates. We also present a fast algorithm for computing the discrete BRT forward operator. Our generalized approach can incorporate both transmission and scatter measurements from multiple source-detector pairs. Transmission measurements resolve low-frequency ambiguity in the joint image estimation problem, while multiple scatter measurements resolve the attenuation image. The benefits of joint estimation, over single-image estimation, vary with problem scaling. Our results quantify these benefits and should inform design of future acquisition systems.

## I. Introduction

**F**OR tomographic imaging applications, scattering is frequently generalized in the extreme: multiple scatter events, or no scatter events. For low-energy (e.g. optical) sources, light typically changes directions multiple times within the media of interest. As the source energy increases, the likelihood of scattering events decreases. Acquisition systems utilizing ionizing radiation are typically designed to ensure most detected photons travel a straight path from the source to the detector. However, there remains an important class of problems where each detected photon is associated with exactly one scatter event: tomography from single-scatter measurements.

To both clarify the broken ray transform and motivate the computational imaging problem, we first define a notional measurement geometry. We illustrate a two-dimensional, single-scatter measurement geometry in Figure 1a. In this case, a pencil-beam source is directed, opposite $\theta_0$, through the media of interest. Along the beam, scattering is observed in two directions $\{\theta_1, \theta_2\}$ for each scatter location $y \in \mathcal{Y}$. We use $\boldsymbol{\theta} = (\theta_s, \theta_d)$ to represent an ordered pair of source and detector directions, respectively. Let $\mathcal{I} := \{\boldsymbol{\theta}_1, \boldsymbol{\theta}_2, \ldots\}$ represent the ordered collection of source-detector pairs for which data are available. For the measurement system depicted in Fig. 1a, we have $\mathcal{I} = \{(\theta_0, \theta_1), (\theta_0, \theta_2)\}$. Using a collimated array of detectors, multiple scatter locations are resolved simultaneously along the source beam. Translating the source and detector arrays as a system allows scanning over all sample points $y \in \mathcal{Y}$. We assume the pairs $(i, y)$ are available for all $(i, y) \in \mathcal{I} \times \mathcal{Y}$. Further, $(i, y)$ uniquely define a broken-ray path, from source to detector, through the scatter location $y$.

Along the path of illumination, attenuating media reduces source intensity exponentially according to Beer's law. This path integral can be described as an operator on the attenuation image. For measurements associated with one scatter event, the path integral is the broken ray transform (BRT) [1]–[3]. The available data is then modeled through a nonlinear relationship between the scatter density image and the attenuation image: a nonlinear function embedding the linear BRT operator. Our computational imaging problem is the joint recovery of these two images: scatter density, $\alpha(y)$; and attenuation, $\mu(x)$.

Broadly speaking, we build on work related to single-scatter measurements from two distinct communities. The mathematics community has largely focused on analytic recovery of the attenuation image: BRT inversion [1]–[7] and related star transforms [8], [9]. In contrast, the material science and physics communities have focused on the scatter density image estimation [10]–[15].

All analytic BRT inversion formulas operate on the log of the measured data. When the log data are finite, analytically the BRT is invertible [5]. With two scatter angles, both the attenuation image and scatter image can be recovered [4]. However, zero scatter yields infinite log data. Recent applications such as luggage scanning [16] require imaging media comprising regions with zero scatter. Additionally, poor

*M. R. Walker II was with the Preston M. Green Department of Electrical and Systems Engineering, Washington University in St. Louis, St. Louis, MO 63130 USA. He is now with Sandia National Laboratories, Albuquerque, NM 87185 (mwalkerii@wustl.edu)

J. A. O'Sullivan is with the Preston M. Green Department of Electrical and Systems Engineering, Washington University in St. Louis, St. Louis, MO 63130 USA (jao@wustl.edu)

J A O'Sullivan was supported by NIH R01 CA 212638

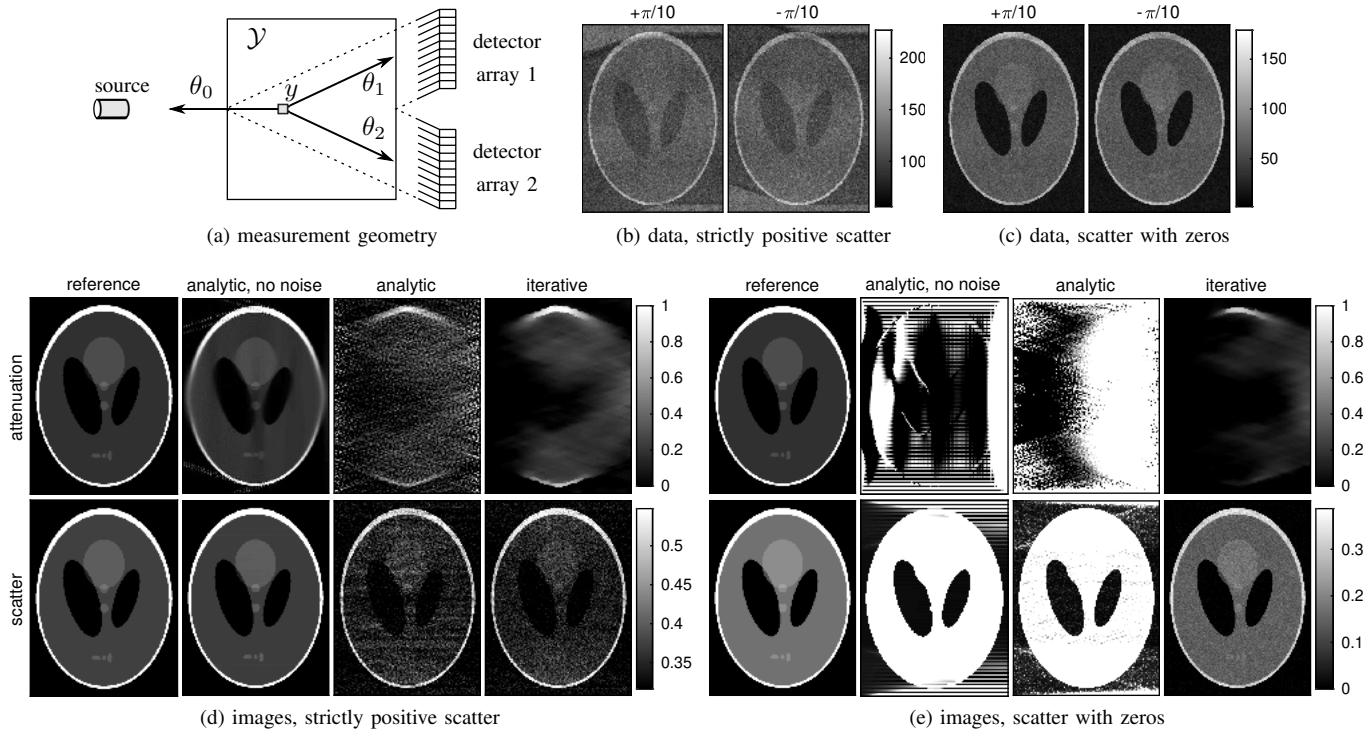

Figure 1. Contrasting analytic and iterative image reconstructions for the single-scatter measurement geometry. The measurement geometry of Fig. 1a comprises a pencil-beam source and two collimated detector arrays. We use $y \in \mathcal{Y}$ to indicate the location of the scatter event. The direction $\theta_0$ indicates the direction of the source from the scatter location $y$. The directions $\theta_i$, $i \in \{1, 2\}$, indicate the observed scatter direction using the $i^{\text{th}}$ detector array. Simulated data for media with strictly positive scatter and scatter with zeros are shown in Fig. 1b and Fig. 1c, respectively, with corresponding images shown in Fig. 1d and Fig. 1e, respectively. In both Fig. 1d and Fig. 1e the columns, from left to right, distinguish: reference images, analytic reconstructions from noise-free data, analytic reconstructions from noisy data, and estimates using our iterative algorithm from noisy data. The results in the third and fourth columns were derived from the same data.

conditioning of the inverse problem [3], [6] is exacerbated by sampling and noise. As shown in Figure 1, applying prior analytic methods to noisy measurements with missing data does not yield useful results.

For scatter density estimation, the attenuation image is often simplified as constant [12], [15] or negligible [11], [13], [14]. Even if the attenuation image is known, exploiting this knowledge remains challenging. Correcting attenuation effects is computationally expensive. Further, when separate acquisition systems are necessary, registration problems appear.

Our goal is to investigate the benefits of joint estimation of scatter and attenuation images from noisy single-scatter measurements. In contrast with prior analytic inversion formulas, our use of stochastic data models is motivated by the low photon counts associated with scatter imaging [10], [11], [13], [14]. Since both scatter density and attenuation are media dependent, joint image estimation could improve object identification [13]. Even if one image is acquired separately, joint estimation can be useful in resolving registration errors. This leaves open questions such as how do errors in the attenuation image affect scatter estimation, and what are the data requirements for joint image estimation?

and its adjoint which are used frequently in iterative reconstruction algorithms. We are also the first to consider joint image estimation from noisy single-scatter measurements with missing data (regions with zero scatter). We apply techniques familiar to the computational imaging community to data models incorporating the BRT operator. Our generalized iterative algorithm can incorporate multiple source locations, scatter angles, and transmission measurements. Our algorithm can be specialized for single-image recovery: scatter image estimation with known attenuation, and attenuation image estimation with known scatter. While alternating image updates, we guarantee monotonic improvement of the penalized log-likelihood. In contrast to prior analytic inversion strategies, our results are significant as strictly positive scatter is not required. For these cases with missing data, we demonstrate ambiguity in the joint image estimation problem. Our generalized algorithm is used to demonstrate the benefits of incorporating transmission measurements for resolving ambiguities in joint image estimation. One projection significantly reduces errors in scatter image estimates. We also demonstrate the benefit of multiple source locations for attenuation estimation which is plagued by poor scaling of the BRT operator.

### A. Main Contributions

This paper has several contributions. We present a computationally efficient algorithm for the discrete BRT operator

### B. Background

*1) Relevant Modalities:* Our interest in single-scatter measurements was motivated by momentum transfer estimation

for luggage scanning applications [16]. However, the single-scatter approximation is reasonable whenever the mean free path is large relative to the media of interest. The list of applicable modalities includes x-ray scatter imaging: fluorescence, which is distinguished as isotropic incoherent scatter; Bragg scatter, distinguished by anisotropic coherent scatter; and Compton scatter, distinguished by anisotropic incoherent scatter [17]. This class of problems also includes optical tomography for optically thin media [1].

Combining data from multiple scatter angles is trivial for media scattering isotropically. This extends easily to x-ray fluorescence imaging, but requires additional conditions for Bragg and Compton scatter imaging [3], [18]. Incorporating three or more source-detector pairs facilitates local analytic reconstruction of the attenuation image with improved reconstruction quality [2]. For coherent scatter imaging applications, scatter images vary with momentum transfer [10]. This restricts combinations of scatter measurements to those with consistent cone angles (i.e. $\theta_s \cdot \theta_d$ constant for all $i \in \mathcal{I}$) [3], [19].

Single-scatter imaging has an analog in positron emission tomography (PET) [2]. The problem of joint estimation of activity and attenuation has received considerable attention [20]–[23]. In particular, the problem was ill-posed until the addition of time-of-flight data (TOF) [21]–[23]. The joint recovery problem for TOF-PET was first solved analytically [22], and then applied to Poisson data models [23], [24]. The additional TOF information limits activity to a segment of the line-of-response. This is analogous to the so-called selected volume around the vertex of the BRT. In this way, TOF-PET can be considered a specialization of single-scatter measurements where $\theta_s = -\theta_d$. It is worth noting, however, that analytic BRT inversion formulas are not useful when the scatter directions are opposite as in PET [8], [19]. Each BRT path has one scatter location by definition, while each line-of-response in PET supports multiple TOF values.

*2) Measurement Geometries:* We consider a simplified measurement geometry comprising a pencil beam source and collimated array of detectors as depicted in Fig. 1a. Image recovery from our measurement geometry has been referred to as selected volume tomography (SVT) [17]. Our specific measurement geometry has also been referred to as the BRT measurement system [2] and a translation-only measurement geometry [3]. Our measurement geometry is distinct from rotational measurement geometries [6], [25]–[28] (related to secondary radiation tomography [17]). Additionally, we distinguish measurement geometries employing coded apertures [11], [13], [14].

Coded aperture measurement geometries share commonality with SVT in that both avoid rotation, and the single-scatter approximation is utilized. However, each measurement is not associated with a unique broken-ray path. Further, the source-detector paths do not all share a common plane thus eluding two-dimensional analysis. Each detector pixel in a coded aperture measurement geometry can be summarized as integrating sparse samples in a high-dimensional space: comprising SVT measurements at multiple scatter angles.

*3) Surrogates in Computational Imaging:* Surrogates have been used in image reconstruction algorithms for many years. The descriptions of approaches includes the EM algorithm itself [29], [30], optimization transfer [31], and minorize maximization [32]. A goal of these techniques is for the surrogate function to be separated as itemizations of parameters of interest. Highly parallelizable algorithms are computationally efficient on modern (GPU) architectures. Additionally, surrogates can be derived for regularized objective functions [33]. In contrast, we derive separate surrogates for the regularization and data fidelity terms before combining the results.

*C. Outline*

The rest of this paper is organized as follows. In Section II we focus on the BRT operator. We first describe the operator analytically in both the spatial and frequency domains. The frequency domain representation highlights several challenges such as poor conditioning and unbounded support. In Section II-B we present a fast implementation of the discrete BRT operator in the frequency domain. Our approach is not obvious due to unbounded support of the analytic BRT data. In Section III we state the computational imaging problem. We start with a statistical data model which motivates our objective function. Rather than attacking the problem directly, joint image reconstruction is addressed by alternating minimization of local surrogate functions. In Section III-B we prove monotonic reduction of the data log-likelihood and convergence to a fixed point. Algorithm implementations are presented in Section III-C. We detail regularized image updates and the incorporation of transmission data. Simulations and results are presented in Section IV. Finally, we discuss special cases and extensions of this work in Section V.

## II. Broken Ray Transform

*A. Analytic Operator*

Before incorporating the BRT in the data model, we first isolate the operator in a purely analytic setting. We consider the transform both in the spatial and frequency domain. This highlights the benefits of our measurement geometry in addition to numerical challenges for reconstruction.

The BRT is the superposition of two improper line integrals sharing a common origin. For an analytic treatment we interpret $x \in \mathbb{R}^2$, and $\mu(x) : \mathbb{R}^2 \to \mathbb{R}_{\geq 0}$. We first define the two-dimensional cone beam transform (CBT) of $\mu$ along the direction $\theta \in S^1$ [3], [34]

$$(C\mu)(x, \theta) := \int_0^\infty \mu(x + t\theta) dt. \quad (1)$$

The CBT is also referred to as the divergent beam transform [9], [35]. The BRT is the superposition of two CBTs

$$(B\mu)(x, \boldsymbol{\theta}) := (C\mu)(x, \theta_s) + (C\mu)(x, \theta_d) \quad (2a)$$

$$= \sum_{\theta \in \boldsymbol{\theta}} \int_0^\infty \mu(x + t\theta) dt. \quad (2b)$$

The summation is over elements of the ordered pair $\boldsymbol{\theta} = (\theta_s, \theta_d)$. The BRT is both linear and shift invariant (LSI). Shift invariance of the BRT is the motivation for indexing the data by scatter location in the measurement geometry.

As an LSI operator, the Fourier transform of BRT data factors as [3]

$$\tilde{b}_{\boldsymbol{\theta}}(w) := \mathcal{F}^2 \left\{ (B\mu)(x, \boldsymbol{\theta}) e^{-j2\pi w \cdot x} \right\} \tag{3a}$$
$$= \tilde{\mu}(w) \tilde{h}_B(w; \boldsymbol{\theta}). \tag{3b}$$

Here $\tilde{\mu}$ and $\tilde{h}$ represent the two-dimensional Fourier transform of the image and data, respectively. The Fourier transform of the CBT operator

$$\tilde{h}_C(w; \theta) := \frac{-1}{j2\pi w \cdot \theta} + \frac{1}{2}\delta(w \cdot \theta) \tag{4}$$

simplifies the Fourier transform of the BRT operator

$$\tilde{h}_B(w; \boldsymbol{\theta}) := \tilde{h}_C(w; \theta_s) + \tilde{h}_C(w; \theta_d) \tag{5a}$$
$$= \frac{jw \cdot (\theta_s + \theta_d)}{2\pi (w \cdot \theta_s)(w \cdot \theta_d)} + \sum_{\theta \in \boldsymbol{\theta}} \frac{1}{2}\delta(w \cdot \theta). \tag{5b}$$

The frequency domain representation of the BRT operator highlights two challenges. First, zeros of (5b) indicate poor scaling in the data. Edges in the image (e.g. $\mu$) along the direction $\theta_s + \theta_d$ are difficult to resolve from BRT data. Second, (5b) is undefined along the lines orthogonal $\theta_s$, and $\theta_d$. These singularities are associated with the unbounded support of the data. Explicitly, BRT data are constant in both directions $-\theta_s$ and $-\theta_d$ beyond the support of the image [3]. Since the BRT data are both unbounded and aperiodic, the data cannot be represented in discrete Fourier space. This is a problem in seeking a computationally efficient discrete BRT operators. While discrete LSI operators typically enjoy computationally efficient implementations leveraging the discrete Fourier transform (DFT), we must first address support of the BRT data in a computationally efficient manor.

### B. Fast Discrete BRT Operators

As a linear operator, the BRT can be implemented directly on sampled data using matrix multiplication. In the following we propose a computationally efficient frequency-domain implementation. For this, we must address the unbounded aperiodic support of the data. Our approach includes applying a new filter function to the image before applying the BRT then truncating the result.

Discrete BRT operators have not been discussed previously beyond a direct implementation. However, previous analytic inversion strategies have addressed discrete Fourier representations of the data. Bounded support of the data can be guaranteed by first applying a filter to the image [3]. The previously proposed filter comprises four delta functions effecting four shifted copies of the original image. Applying the BRT to the filtered image results in four shifted copies of the desired data. The filter design ensures data copies coherently cancel outside a bounded region of support (a parallelogram). The filter is parameterized by spreading lengths which determine overlap between the data copies. In general, small scatter angles require larger spreading lengths to avoid overlap. Equivalently, padding of the DFT is required to avoid aliasing. Padding is not a problem for an analytic inversion formula. However, for our purposes padding undermines the computational efficiency of the forward operator.

Instead of applying the BRT to a filtered image with bounded support, we decompose the BRT as two CBTs. We apply separate filters to the inputs of the two CBTs and combine the results. This effects discrete BRT data with symmetric boundaries. The benefit is a reduction in the required DFT padding yielding computationally efficient implementations.

Symmetric boundaries of the data requires assumptions on sampling and scatter angles. First we assume the scatter location correspond to image samples $\mathcal{Y} = \mathcal{X}$. We assume a uniform orthogonal sampling lattice with $L_2 \times L_1$ samples with sample spacing $\Delta_2, \Delta_1 \in \mathbb{R}_{>0}$. We require at least one sample axis to be aligned with a BRT direction. The alignment requirement is not critical, but it simplifies the required padding and filter definitions.

Without loss of generality, we assume $\theta_s$ is aligned with the horizontal sampling axis. Our filter function is parameterized by spreading factors associated with the two BRT directions

$$a_s = L_1 \Delta_1, \quad a_d = \frac{a_s}{|\theta_s \cdot \theta_d|}. \tag{6}$$

We employ two filter functions, one for each CBT

$$m_s(x) = \delta(x) - \delta(x + 3a_s\theta_s + a_d\theta_d)$$
$$\quad - \delta(x + 2a_s\theta_s) + \delta(x + 2a_s\theta_s + a_d\theta_d) \tag{7}$$
$$m_d(x) = \delta(x) - \delta(x + a_d\theta_d)$$
$$\quad - \delta(x + 2a_s\theta_s) + \delta(x + 2a_s\theta_s + a_d\theta_d). \tag{8}$$

These functions differ only by the second right-hand terms. Their periodic extensions are equivalent when repeated in the direction $\theta_s$ with period $3a_s$. The two-dimensional Fourier transforms of these functions read

$$\widetilde{m}_s(w) = -j2\sin(2\pi a_s w \cdot \theta_s) e^{j2\pi a_s w \cdot \theta_s}$$
$$\quad - j2\sin(\pi a_s w \cdot \theta_s) e^{j5\pi a_s w \cdot \theta_s} e^{j2\pi a_d w \cdot \theta_d} \tag{9}$$
$$\widetilde{m}_d(w) = -4\sin(2\pi a_s w \cdot \theta_s) \sin(\pi a_d w \cdot \theta_d)$$
$$\quad \times e^{j2\pi a_s w \cdot \theta_s} e^{j\pi a_d w \cdot \theta_d}. \tag{10}$$

To obtain the filtered BRT, we must apply the filter function to the corresponding CBT and sum the results. Making use of (4), we have

$$\tilde{h}_{\text{FBRT}}(w) := \widetilde{m}_s(w) \tilde{h}_C(w; \theta_s) + \widetilde{m}_d(w) \tilde{h}_C(w; \theta_d) \tag{11a}$$
$$= 2a_s \operatorname{sinc}(2a_s w \cdot \theta_s) e^{j2\pi a_s w \cdot \theta_s}$$
$$\quad + a_s \operatorname{sinc}(a_s w \cdot \theta_s) e^{j5\pi a_s w \cdot \theta_s} e^{j2\pi a_d w \cdot \theta_d}$$
$$\quad + \frac{2a_d}{j} \sin(2\pi a_s w \cdot \theta_s) \operatorname{sinc}(a_d w \cdot \theta_d)$$
$$\quad \times e^{j2\pi a_s w \cdot \theta_s} e^{j\pi a_d w \cdot \theta_d}. \tag{11b}$$

In (11b) we omit the delta functions sampling (9) and (10) where they are zero. In contrast to (5b), the expression (11b) is well-defined everywhere.

To accommodate the increased extent of filtered data, we must pad the original image size

$$N_1 = 3L_1, \quad N_2 = L_2 + \lceil a_d \|\theta_s \times \theta_d\| / \Delta_2 \rceil. \tag{12}$$

The ceiling operation is indicated using $\lceil \cdot \rceil$. The horizontal padding is selected precisely to support nonzero symmetric boundaries. The vertical padding is only included to avoid aliasing. Fig. 2 illustrates our operator on padded data.

Our implementation of the forward operator first requires sampling the image on a rectangular grid. We then compute the two-dimensional DFT using zero-padding determined by (6) and (12). The result is multiplied element-wise by (11b) sampled at the frequencies determined by the padded DFT lengths. Finally, the inverse DFT is applied and the result is truncated to the original $L_2 \times L_1$ image size. For a more detailed description, see [19]. We note the backward BRT operator is computed by conjugating each sample of (11b).

## III. COMPUTATIONAL IMAGING PROBLEM

### A. Data Model

Our objective is the recovery of two images, attenuation and scatter density, from a single dataset. For clarity we use separate discretizations of the images using $x \in \mathcal{X}$ and $y \in \mathcal{Y}$ to index the set of attenuation and scatter image voxels, respectively. Indexing the data by the scatter location, we use the same index for both the data and the scatter image. Let $\mathcal{A} \coloneqq \{\boldsymbol{\mu} : \mu(x) \in \mathbb{R}_{\geq 0}\}$ and $\mathcal{S} \coloneqq \{\boldsymbol{\alpha} : \alpha(y) \in [0,1]\}$ represent the set of possible attenuation and scatter images, respectively. Let $\mathbf{d}$ represent the available data where $d_i(y) \in \mathbb{R}_{\geq 0}$ for each $(i, y) \in \mathcal{I} \times \mathcal{Y}$.

We model the data $d_i(y)$ as Poisson distributed with mean

$$g_i(y : \boldsymbol{\alpha}, \boldsymbol{\mu}) \coloneqq \beta_i(y) + I_0(y)\alpha(y) \exp\left(-\sum_{x \in \mathcal{X}} h_i(y|x)\mu(x)\right). \quad (13)$$

Here we use $I_0(y) > 0$ and $\beta(y) \geq 0$ to represent the known source intensity and background counts, respectively. The exponential term is a numeric approximation of Beer's law along the source-detector path through the attenuation image. For single-scatter measurements, $h_i$ represents the discrete BRT of the attenuation image

$$(B\mu)(x, \boldsymbol{\theta}_i) \approx \sum_{x \in \mathcal{X}} h_i(y|x)\mu(x). \quad (14)$$

This is an approximation to the analytic BRT (2b). The subscript $i$ appears on both sides and indexes source-receiver pairs. We assume both the image and forward transform are finite everywhere.

The log-likelihood function of the data, parameterized by $\boldsymbol{\alpha}, \boldsymbol{\mu}$, is

$$l(\mathbf{d} : \boldsymbol{\alpha}, \boldsymbol{\mu}) \coloneqq \sum_{i \in \mathcal{I}} \sum_{y \in \mathcal{Y}} d_i(y) \ln g_i(y : \boldsymbol{\alpha}, \boldsymbol{\mu}) - g_i(y : \boldsymbol{\alpha}, \boldsymbol{\mu}) \quad (15)$$

excluding constant terms of the data alone. Maximizing the log-likelihood is equivalent to minimizing the Csiszár I-divergence between the data and the mean

$$I(\mathbf{d} \parallel \mathbf{g}(\boldsymbol{\alpha}, \boldsymbol{\mu})) \coloneqq \sum_{i \in \mathcal{I}} \sum_{y \in \mathcal{Y}} d_i(y) \ln \frac{d_i(y)}{g_i(y : \boldsymbol{\alpha}, \boldsymbol{\mu})} - d_i(y) + g_i(y : \boldsymbol{\alpha}, \boldsymbol{\mu}). \quad (16)$$

Here we use $\mathbf{g}(\boldsymbol{\alpha}, \boldsymbol{\mu}) \coloneqq \{g_i(y : \boldsymbol{\alpha}, \boldsymbol{\mu}) : i \in \mathcal{I}\}$. This form is a generalization of the Kullback-Leibler divergence [36].

### B. Objective Functions and Surrogates

Joint image recovery from (16) is ill-posed due to conditioning of the BRT forward operator [3] in the exponent (13), noise, and scaling. To improve conditioning of this problem we incorporate two regularization terms in the objective function

$$J(\boldsymbol{\alpha}, \boldsymbol{\mu}) \coloneqq I(\mathbf{d} \parallel \mathbf{g}(\boldsymbol{\alpha}, \boldsymbol{\mu})) + \lambda_\alpha R(\boldsymbol{\alpha}) + \lambda_\mu R(\boldsymbol{\mu}). \quad (17)$$

Here $R$ is a convex regularization function (further conditions given in Appendix C). The scalars $\lambda_\alpha$ and $\lambda_\mu$ emphasize regularization of the corresponding images.

Direct minimization of (17) remains difficult due to its high dimensionality and interdependence of the image pixels. To make the problem tractable, we employ two techniques. First, we use separable surrogate functions for the terms in (17). Separability here means the gradients separate as functions of single image samples. The ensuing algorithm is highly parallelizable, allowing each pixel update to be computed in parallel. Second, we employ alternating updates between the scatter and attenuation images. The use of surrogate functions guarantees monotonic reduction in the objective while alternating image updates.

We use a surrogate for the data fidelity term

$$\overline{D}(\boldsymbol{\mu} : \hat{\boldsymbol{\alpha}}, \hat{\boldsymbol{\mu}}) \geq I(\mathbf{d} \parallel \mathbf{g}(\hat{\boldsymbol{\alpha}}, \boldsymbol{\mu})), \quad \forall \boldsymbol{\mu} \in \mathcal{A} \quad (18)$$

$$\overline{D}(\hat{\boldsymbol{\mu}} : \hat{\boldsymbol{\alpha}}, \hat{\boldsymbol{\mu}}) = I(\mathbf{d} \parallel \mathbf{g}(\hat{\boldsymbol{\alpha}}, \hat{\boldsymbol{\mu}})), \quad (19)$$

which is given by (68) and derived in Appendix B.

Additionally, we consider a surrogate for the regularization term

$$\overline{R}(\boldsymbol{\mu} : \hat{\boldsymbol{\mu}}) \geq R(\boldsymbol{\mu}), \quad \forall \boldsymbol{\mu} \in \mathcal{A} \quad (20)$$

$$\overline{R}(\hat{\boldsymbol{\mu}} : \hat{\boldsymbol{\mu}}) = R(\hat{\boldsymbol{\mu}}), \quad (21)$$

which is given by (82) and derived in Appendix C.

Making use of these surrogate functions, we define two objectives

$$\overline{J}_\alpha(\boldsymbol{\alpha} : \hat{\boldsymbol{\alpha}}, \hat{\boldsymbol{\mu}}) \coloneqq I(\mathbf{d} \parallel \mathbf{g}(\boldsymbol{\alpha}, \hat{\boldsymbol{\mu}})) + \lambda_\alpha \overline{R}(\boldsymbol{\alpha} : \hat{\boldsymbol{\alpha}}) \quad (22)$$

$$\overline{J}_\mu(\boldsymbol{\mu} : \hat{\boldsymbol{\alpha}}, \hat{\boldsymbol{\mu}}) \coloneqq \overline{D}(\boldsymbol{\mu} : \hat{\boldsymbol{\alpha}}, \hat{\boldsymbol{\mu}}) + \lambda_\mu \overline{R}(\boldsymbol{\mu} : \hat{\boldsymbol{\mu}}), \quad (23)$$

which lead to an iterative update algorithm

$$\boldsymbol{\alpha}^{(k+1)} = \arg\min_{\boldsymbol{\alpha} \in \mathcal{S}} \overline{J}_\alpha(\boldsymbol{\alpha} : \boldsymbol{\alpha}^{(k)}, \boldsymbol{\mu}^{(k)}) \quad (24)$$

$$\boldsymbol{\mu}^{(k+1)} = \arg\min_{\boldsymbol{\mu} \in \mathcal{S}} \overline{J}_\mu(\boldsymbol{\mu} : \boldsymbol{\alpha}^{(k+1)}, \boldsymbol{\mu}^{(k)}). \quad (25)$$

This approach guarantees monotonic reduction of the regularized objective and convergence to a local minimum.

**Lemma 1.** *Monotonic reduction of $J(\boldsymbol{\alpha}, \boldsymbol{\mu})$ is guaranteed whenever a local objective,* (22) *or* (23), *is reduced.*

*Proof.* Combining the definitions (17) and (22), and making use of (20), we have

$$J(\boldsymbol{\alpha}, \hat{\boldsymbol{\mu}}) - \lambda_\mu R(\hat{\boldsymbol{\mu}}) \leq \overline{J}_\alpha(\boldsymbol{\alpha} : \hat{\boldsymbol{\alpha}}, \hat{\boldsymbol{\mu}}). \quad (26)$$

According to (21), we have equality when $\boldsymbol{\alpha} = \hat{\boldsymbol{\alpha}}$, such that

$$J(\boldsymbol{\alpha}, \hat{\boldsymbol{\mu}}) - J(\hat{\boldsymbol{\alpha}}, \hat{\boldsymbol{\mu}}) \leq \overline{J}_\alpha(\boldsymbol{\alpha} : \hat{\boldsymbol{\alpha}}, \hat{\boldsymbol{\mu}}) - \overline{J}_\alpha(\hat{\boldsymbol{\alpha}} : \hat{\boldsymbol{\alpha}}, \hat{\boldsymbol{\mu}}). \quad (27)$$

Therefore, any $\boldsymbol{\alpha}$ reducing $\overline{J}_\alpha$ guarantees a reduction in $J$. Further, the improvement in the objective is bounded by the improvement to the local surrogate. The same can be shown for any $\boldsymbol{\mu}$ reducing $\overline{J}_\mu$. □

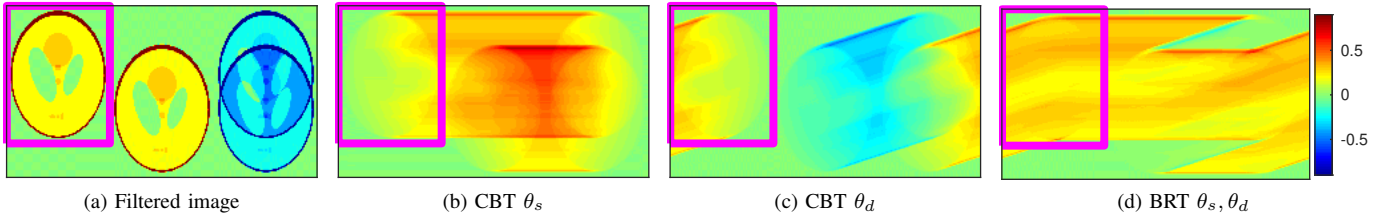

(a) Filtered image  (b) CBT $\theta_s$  (c) CBT $\theta_d$  (d) BRT $\theta_s, \theta_d$

Figure 2. Filtered images effecting periodic CBT and BRT data. One tile of the periodic filtered image is shown in Figure 2a. CBT data associated with the directions $\theta_s$ and $\theta_d$ are shown in Figures 2b and 2c, respectively. Summing these results, we obtain the periodic BRT data shown in Fig. 2d. The magenta rectangle indicates the support of the original image and cropping boundaries to obtain the truncated BRT data of interest.

Applying Lemma 1, alternating updates ensure
$$J(\boldsymbol{\alpha}^{(k)}, \boldsymbol{\mu}^{(k)}) \geq J(\boldsymbol{\alpha}^{(k+1)}, \boldsymbol{\mu}^{(k)}) \geq J(\boldsymbol{\alpha}^{(k+1)}, \boldsymbol{\mu}^{(k+1)}). \quad (28)$$
Iterative updates result in a sequence of costs that are monotonically decreasing and bounded from below, since $J(\boldsymbol{\alpha}, \boldsymbol{\mu}) \geq 0$. Convergence of this sequence is guaranteed.

**Lemma 2.** *Convergence of* (24) *and* (25) *implies a fixed point.*

*Proof.* The divergence (16) is convex with respect to $\boldsymbol{\alpha}$ (see Appendix A), and $\overline{R}$ is strictly convex (see Appendix C). Therefore, the local surrogate $\overline{J}_\alpha$ is strictly convex over $\boldsymbol{\alpha} \in \mathcal{S}$, with a unique minimizer $\boldsymbol{\alpha}^*$. When (24) does not improve the objective, we have
$$\overline{J}_\alpha(\boldsymbol{\alpha}^* : \boldsymbol{\alpha}^{(k)}, \boldsymbol{\mu}^{(k)}) = \overline{J}_\alpha(\boldsymbol{\alpha}^{(k)} : \boldsymbol{\alpha}^{(k)}, \boldsymbol{\mu}^{(k)})$$
$$\implies \boldsymbol{\alpha}^* = \boldsymbol{\alpha}^{(k)} \quad (29)$$
such that $\boldsymbol{\alpha}^{(k)}$ is a fixed point. The same can be shown for $\overline{J}_\mu$. □

Applying Lemma 2, equality in (28) implies $\boldsymbol{\alpha}^{(k+1)} = \boldsymbol{\alpha}^{(k)}$ and $\boldsymbol{\mu}^{(k+1)} = \boldsymbol{\mu}^{(k)}$.

### C. Algorithms

Our iterative joint estimation approach is a two-step process summarized by Algorithm 1. The first step requires computation of the forward BRT, and the second requires computation of the backward BRT.

---
**Algorithm 1** JOINTESTIMATE: Iterative algorithm for joint image estimation
---
**Input:** $\boldsymbol{\alpha}^{(0)}, \boldsymbol{\mu}^{(0)}$
1: **for** $k = 0, 1, 2, \ldots$ **do**
2: $\quad \boldsymbol{\alpha}^{(k+1)} = \text{SCATTERUPDATE}\left(\boldsymbol{\alpha}^{(k)}, \boldsymbol{\mu}^{(k)}\right)$
3: $\quad \boldsymbol{\mu}^{(k+1)} = \text{ATTENUATIONUPDATE}\left(\boldsymbol{\alpha}^{(k+1)}, \boldsymbol{\mu}^{(k)}\right)$
4: **end for**
---

*1) Regularized Scatter Update:* The objective function (22) is strictly convex with respect to $\boldsymbol{\alpha}$ over $\mathcal{S}$ (see Appendices A and C). Therefore, each $\alpha(y)$ is either 0 or the solution to an unconstrained minimization problem. Nonzero voxels are then determined by setting the gradient of (22) equal to zero and solving for $\alpha(y)$. Expanding (22), we have
$$\frac{\partial \overline{J}_\alpha(\boldsymbol{\alpha} : \boldsymbol{\alpha}^{(k)}, \boldsymbol{\mu}^{(k)})}{\partial \alpha(y)} = \frac{\partial I(\mathbf{d} \parallel \mathbf{g}(\boldsymbol{\alpha}, \boldsymbol{\mu}^{(k)}))}{\partial \alpha(y)}$$
$$+ \lambda_\alpha \frac{\partial \overline{R}(\boldsymbol{\alpha} : \boldsymbol{\alpha}^{(k)})}{\partial \alpha(y)}. \quad (30)$$

The first and second terms on the right-hand side are given by (43) and (83), respectively. Expanding these terms, we find the gradient is separable with respect to $\alpha(y)$. Each $\alpha^{(k+1)}(y)$ can be determined in parallel solving
$$0 = \sum_{i \in \mathcal{I}} \left( \dot{g}_i(y : \boldsymbol{\mu}^{(k)}) - \frac{d_i(y) \dot{g}_i(y : \boldsymbol{\mu}^{(k)})}{\alpha(y) \dot{g}_i(y : \boldsymbol{\mu}^{(k)}) + \beta_i(y)} \right)$$
$$+ \lambda_\alpha \left( c_1(y : \boldsymbol{\alpha}^{(k)}) + 2c_2(y : \boldsymbol{\alpha}^{(k)}) \left( \alpha(y) - \alpha^{(k)}(y) \right) \right). \quad (31)$$

Here $\dot{g}$ is given by (42). For each $i \in \mathcal{I}$, computing $\dot{g}_i$ requires computing the forward BRT of $\boldsymbol{\mu}^{(k)}$. The functions $c_1$ and $c_2$ refer to (80) and (81), respectively. These have been re-appropriated for use with the scatter image and depend on the previous estimate $\boldsymbol{\alpha}^{(k)}$. This process is described in Algorithm 2.

In (31), at most one solution $\alpha(y) \geq 0$ exists since (22) is strictly convex over $\boldsymbol{\alpha} \in \mathcal{A}$. The first term in (31) is monotonic and the second term is a line. When the cardinality $\mathcal{I}$ is small, we derive the polynomial form and solve for the largest root. For $|\mathcal{I}| > 2$, we found Newton's method for root-finding computationally efficient.

---
**Algorithm 2** SCATTERUPDATE: Single update of the scatter image.
---
**Input:** $\hat{\boldsymbol{\alpha}}, \hat{\boldsymbol{\mu}}$
**Output:** $\boldsymbol{\alpha}$
1: **for each** scatter angle $i$ **do**
2: $\quad b_i(y) = \sum_{x \in \mathcal{X}} h_i(y|x) \hat{\boldsymbol{\mu}}(x)$ ▷ Forward BRT
3: $\quad \dot{g}_i(y) = I_0(y) \exp(-b_i(y))$
4: **end for**
5: Compute $c_1(y : \hat{\boldsymbol{\alpha}})$ using Eq. (80)
6: Compute $c_2(y : \hat{\boldsymbol{\alpha}})$ using Eq. (81)
7: **for each** point $y$ **do**
8: $\quad$ Compute $\alpha(y)$ by solving Eq. (31)
9: **end for**
---

*2) Regularized Attenuation Update:* The objective function (23) is strictly convex with respect to $\boldsymbol{\mu}$ over $\mathcal{A}$ (see Appendices B and C). Therefore, each $\mu(x)$ is either 0 or the solution to an unconstrained minimization problem. Nonzero voxels are then determined by setting the gradient of (23) equal to zero and solving for $\mu(x)$. Expanding (23), we have

$$\frac{\partial \overline{J}_\mu(\boldsymbol{\mu} : \boldsymbol{\alpha}^{(k+1)}, \boldsymbol{\mu}^{(k)})}{\partial \mu(x)} = \frac{\partial \overline{D}(\boldsymbol{\mu} : \boldsymbol{\alpha}^{(k+1)}, \boldsymbol{\mu}^{(k)})}{\partial \mu(x)} + \lambda_\mu \frac{\partial \overline{R}(\boldsymbol{\mu} : \boldsymbol{\mu}^{(k)})}{\partial \mu(x)}. \quad (32)$$

The first and second terms on the right-hand side are given by (69) and (83), respectively. Expanding these terms, we find the gradient is separable with respect to $\mu(x)$. Each $\mu^{(k+1)}(x)$ can be determined in parallel solving

$$\begin{aligned}
0 = &\, b_1(x : \boldsymbol{\alpha}^{(k+1)}, \boldsymbol{\mu}^{(k)}) \\
&- b_2(x : \boldsymbol{\alpha}^{(k+1)}, \boldsymbol{\mu}^{(k)}) \exp\Big(-Z_0\big(\mu(x) - \mu^{(k)}(x)\big)\Big) \\
&+ \lambda_\mu \Big(c_1(x : \boldsymbol{\mu}^{(k)}) + 2c_2(x : \boldsymbol{\mu}^{(k)})\big(\mu(x) - \mu^{(k)}(x)\big)\Big).
\end{aligned} \quad (33)$$

The functions $b_1$ and $b_2$ are given by (66) and (67), respectively. Here we emphasize their dependence on prior estimates $\boldsymbol{\alpha}^{(k+1)}$ and $\boldsymbol{\mu}^{(k)}$. Prior estimates are used to compute $\hat{q}_i(y, 1)$ using (46), which determine $\hat{p}_i(y, 1)$ using (49b). The adjoint BRT is required to compute $b_1$ and $b_2$ from $\hat{p}_i(y, 1)$ and $\hat{q}_i(y, 1)$, respectively. This process is described in Algorithm 3.

Similar to (31), the solution to (33) lies at the intersection of a monotonic function and a straight line. Again, we use Newton's method for root-finding.

---

**Algorithm 3** ATTENUATIONUPDATE: Single update of the attenuation image.

---
**Input:** $\hat{\boldsymbol{\alpha}}, \hat{\boldsymbol{\mu}}$
**Output:** $\boldsymbol{\mu}$
1: $\mathbf{b}_1, \mathbf{b}_2 \leftarrow \mathbf{0}$
2: **for each** scatter angle $i$ **do**
3:     Compute $\hat{q}_i(y)$ using Eq. (46)
4:     Compute $\hat{p}_i(y)$ using Eq. (49b)
5:     $b_1(x) \mathrel{+}= \sum_{y \in \mathcal{Y}} h_i(y|x) \hat{p}_i(y)$     ▷ backward BRT
6:     $b_2(x) \mathrel{+}= \sum_{y \in \mathcal{Y}} h_i(y|x) \hat{q}_i(y)$
7: **end for**
8: Compute $c_1(x : \hat{\boldsymbol{\mu}})$ using Eq. (80)
9: Compute $c_2(x : \hat{\boldsymbol{\mu}})$ using Eq. (81)
10: **for each** point $x$ **do**
11:     Compute $\mu(x)$ by solving Eq. (33)
12: **end for**

---

*3) Incorporating Transmission Measurements:* Poor conditioning of the BRT, and missing data due to regions with no scatter, complicate joint image estimation. Depending on acquisition system design, it may be efficient to acquire single-aspect transmission measurements with scatter measurements simultaneously. For example, a single collimated detector could be placed opposite the source and measuring attenuation along straight-line paths. By introducing some notational changes, limited-aspect transmission measurements can be incorporated without changing previous image update expressions.

First we distinguish a subset of source-detector pairs as transmission measurements

$$\mathcal{I}_X := \{\boldsymbol{\theta} \in \mathcal{I} : \theta_s = -\theta_d\}. \quad (34)$$

Transmission measurements are associated with a reduced set of samples: one measurement per line of response. For each source-detector pair, we define $\mathcal{Y}_i \subseteq \mathcal{Y}$. For $\boldsymbol{\theta}_i \notin \mathcal{I}_X$, we retain $\mathcal{Y}_i = \mathcal{Y}$. We assign $d_i(y) = 0$ for all $y \notin \mathcal{Y}_i$. Here the choice of $\mathcal{Y}_i$ is particularly important for indexing known values $I_0(y)$. For transmission measurements, $\mathcal{Y}_i$ may simply index detector locations for all $i \in \mathcal{I}_X$.

The model for mean data counts must be expanded to accommodate transmission measurements. Using $\mathcal{I}_X$ and $\mathcal{Y}_i$, we augment (13) as

$$g_i(y : \boldsymbol{\alpha}, \boldsymbol{\mu}) := \begin{cases} 0, \forall y \notin \mathcal{Y}_i; \\ \beta_i(y) + I_0(y) \exp\left(-\sum_{x \in \mathcal{X}} h_i(y|x) \mu(x)\right), y \in \mathcal{Y}_i, i \in \mathcal{I}_X; \\ \beta_i(y) + I_0(y) \alpha(y) \exp\left(-\sum_{x \in \mathcal{X}} h_i(y|x) \mu(x)\right), i \notin \mathcal{I}_X \end{cases}. \quad (35)$$

The first two cases apply to transmission measurements and do not include $\boldsymbol{\alpha}$. The third case addresses scatter measurements and is equivalent to (13).

The scatter update expression (31) requires no changes accounting for transmission measurements. All additional terms in (16) due to transmission measurements are constant with respect to $\boldsymbol{\alpha}$. However, for tracking reduction in the objective (17), it is important to incorporate transmission measurements and (35) when computing (16).

Additionally, transmission measurements do not change the form of the attenuation update (33). The distinction between transmission and scatter measurements are important when computing $b_2(x)$ given by (67). In particular, $q_i(y; 1)$ should be computed using $\boldsymbol{\alpha} = 1$ for $i \in \mathcal{I}_X$ according to (35). However, these distinctions are eclipsed by $b_2(x)$ leaving (33) unchanged.

## IV. RESULTS

### A. Simulation Setting

Our results are focused on demonstrating the poor conditioning of the joint estimation problem, and the benefits of our generalized iterative approach. Prior analytic approaches are reasonable for recovering attenuation from media with scatter everywhere and high signal to noise ratios. However, missing data and low photon counts associated with single-scatter measurements complicate reconstruction of both images. For media with regions of zero scatter, transmission measurements resolve low-frequency errors in scatter image

Table I
SIMULATION PARAMETERS AND HYPERPARAMETERS

| figure | 1 | 3 | 4 | 5 |
|---|---|---|---|---|
| number of sources | 1 | 1 | 1-2 | 2-8 |
| transmission data | none | none | varies | Hx, Vx |
| scatter angle | $\pm\pi/10$ | $\pm\pi/10$ | $\pi/10$ | varies |
| $\alpha(y)$ | $>0, \geq 0$ | $>0, \geq 0$ | $\geq 0$ | $\geq 0$ |
| $\mu(x)$ | S.L. | Rect. | S.L. | S.L. |
| max $\mu$ | 1 | 3 | 10 | 1 |
| $I_0$ | 350 | 1000 | 1000 | 350 |
| $\beta_0$ | 17.5 | 50 | 50 | 17.5 |
| noise-free | T/F | T | T | F |
| $\lambda_\alpha, \lambda_\mu$ | 2e-3 | 0 | 0 | 2e-3 |

estimation. Still, multiple source locations may be necessary to resolve attenuation images.

In all cases, we simulate data from analytic shapes from which the BRTs are determined analytically. The data are determined using (13) where the summation in the exponent is replaced with samples of the analytic BRT. All data and images are uniformly sampled in orthogonal directions over a rectangular region 300 by 400 pixels.

For the attenuation image, we use two separate phantoms: the modified Shepp-Logan phantom [37], [38], and a binary rectangle. The support of the Shepp-Logan phantom is covered by a 1.5 by 2 (unitless) rectangle. The phantom values $\mu(x) \in [0,1]$ yields maximum BRT values around 0.5, and multiplicative total attenuation terms between 0.6 and 1. For the simulation results shown in Fig. 3 and Fig. 4, we apply a scale factor to the attenuation image. This increases the dynamic range of attenuation effects. In Fig. 3, we use a binary rectangular phantom for the attenuation image that is 1 by 1.5 (unitless).

For the scatter image, we apply one of two nonlinear operators to the attenuation image (before scaling)

$$\alpha_{>0}(y) := \sqrt{0.1 + 0.2\mu(y)} \qquad (36)$$
$$\alpha_{\geq 0}(y) := \sqrt{0.15\mu(y)}. \qquad (37)$$

These transforms ensure transitions in the attenuation image are associated with transitions in the scatter images, which is characteristic of inhomogeneities in the imaging media.

Simulation parameters and algorithm hyperparameters utilized in our results are listed in Table I and distinguished per figure. The row max $\mu$ indicates scaling applied to the attenuation image. The last row indicates whether regularization was utilized, and we set $\lambda_\alpha = \lambda_\mu$ in all cases.

### B. Contrasting Analytic Results

Using analytic BRT inversion strategies to recover images with missing data requires some modification. First, to obtain strictly positive data we apply a threshold

$$\bar{d}_i(y) := \max(d_i(x) - \beta_i(y), d_0). \qquad (38)$$

Assuming counting detectors, we use integer-valued $d_0 = 1$. We recover BRT data using

$$\hat{b}_i(x) := -\ln(\bar{d}_i(y)) + \ln I_0. \qquad (39)$$

From $\hat{b}_i$ we use the analytic inversion for the modified BRT [3] to recover $\hat{\mu}$. We assume two symmetric scatter angles and a common source. Using $\hat{\mu}$, we approximate the scatter image by averaging the thresholded data

$$\hat{\alpha}(y) := \frac{1}{|\mathcal{I}|} \sum_{i \in \mathcal{I}} \left[ \frac{\bar{d}_i(y)}{I_0(y)} \exp\left( \sum_{x \in \mathcal{X}} h_i(y|x)\hat{\mu}(x) \right) \right]. \qquad (40)$$

We contrast analytic and iterative image reconstructions in Fig. 1d and Fig. 1e for simulated scattering media defined by (36) and (37). As demonstrated in Fig. 1d, analytic reconstructions exhibit few artifacts for media with strictly positive scatter and noise-free data. However, performance degrades when the same formula is applied to Poisson distributed data. This is particularly true for the attenuation image. Our approach improves attenuation image reconstruction, although recovery exhibits significant blurring. The scatter image is noisy and includes some low frequency errors. In Fig. 1e we demonstrate the effects of regions with zero scatter. This case constitutes a misapplication of analytic inversion formulas as regions with zero scatter are assumed obscured by regions of high attenuation. In contrast, missing data does not dramatically affect reconstruction quality using our iterative approach. Qualitatively, the scatter image estimate improves while the attenuation image degrades. For low attenuating media with zero-scatter regions, however, the data resemble a scaled version of the scatter image (c.f. Fig. 1c). It is reasonable to expect good scatter image recovery.

### C. Joint Estimation Ambiguity

Analyzing noise-free data from a simple phantom highlights some limitations of joint image recovery from two BRT datasets. We consider a rectangular phantom interrogated by a single source with symmetric scatter angles $\pm\pi/10$ excluding transmission data. Results from noise-free data and no regularization are shown in Fig. 3. In Fig. 3a, scatter is exhibited everywhere, while scatter is limited to the support of the attenuation image in Fig. 3b. In Fig. 3a, attenuation image recovery does not resolve the vertical edges. This is due to zeros in the forward operators and can be mitigated with regularization. The support of the estimated scatter image agrees with the true scatter image, although low frequency errors exists. These errors appear similar to cupping artifacts associated with beam hardening although our simulations do not assume energy sensitive detectors. Instead these errors in the scatter images are coupled with errors in the attenuation image. In Fig. 3b, attenuation image estimates demonstrate significant errors, particularly in regions with missing data. The low frequency errors in the scatter image are more significant. The mere existence of these errors in the noise-free case imply additional data are necessary.

### D. Transmission Measurements

Incorporating transmission measurements resolves low-frequency errors in the scatter images. The benefits are demonstrated in Fig. 4. In particular, incorporating transmission measurements along a single source direction has a significant impact. However, this may degrade attenuation image estimation using some performance metrics (see second column of Fig. 4). Incorporating two orthogonal source directions with transmission measurements improves attenuation image reconstruction. This again follows from the zeros of the BRT operator.

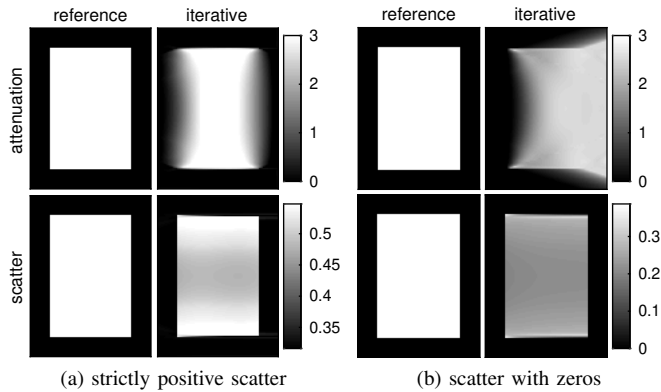

Figure 3. Unregularized reconstructions for a rectangular phantom from noise-free data. Columns distinguish the reference image from our estimates. Rows distinguish attenuation from scatter images. Between Fig. 3a and Fig. 3b the reference scatter image and colorscale change.

### E. Multiple Sources

For small scatter angles, many of the spatial frequencies attenuated by the BRT operator for $\theta_0, \theta_1$ are also attenuated for $\theta_0, \theta_2$ thwarting recovery of these spectral components. However, rotating the source-detector pair reduces the extent of frequencies attenuated by both BRT operators. Multiple source locations are not particularly helpful for scatter image estimation. However, these additional data help resolve the attenuation image which is especially challenging for small scatter angles. These results are demonstrated with Fig. 5.

### F. Processing Time

To demonstrate the computational efficiency of our approach, with respect to problem size, we report iteration and operator processing times in Table II and Table III, respectively. All computations have been preformed in MATLAB on a Late 2016 MacBook Pro (2.9 GHz Quad-Core i7) parallelized among 4 workers with no GPU support. We expect significant reduction in processing times for highly parallelizable systems (GPUs). However, our results quantify effects of problem size on processing burden. In particular, additional scatter angles do not increase processing time multiplicatively. Our Fourier operator implementation significantly improves processing time over a direct (sparse) matrix implementation.

### V. DISCUSSION

Use of the algorithms described here are motivated by two considerations: 1, scatter is zero over significant regions; and 2, data are noisy. For such cases, our simulation results demonstrate significant improvement in reconstructed image quality in comparison with prior analytic reconstruction methods.

Table II
ALGORITHM PROCESSING TIMES IN SECONDS

| | scatter update | | | attenuation update | | |
|---|---|---|---|---|---|---|
| pixels | $|\mathcal{I}|=2$ | 4 | 8 | $|\mathcal{I}|=2$ | 4 | 8 |
| 30000 | 0.17 | 0.20 | 0.19 | 0.24 | 0.27 | 0.34 |
| 120000 | 0.38 | 0.53 | 0.62 | 0.73 | 0.95 | 1.30 |
| 480000 | 1.41 | 2.02 | 2.19 | 2.85 | 3.50 | 4.45 |

Table III
OPERATOR PROCESSING TIMES IN SECONDS

| | direct | | | Fourier | | |
|---|---|---|---|---|---|---|
| pixels | setup | forward | back | setup | forward | back |
| 30K | 9.3e+0 | 7.6e-3 | 3.6e-3 | 3.6e-2 | 4.3e-3 | 4.3e-3 |
| 120K | 3.8e+1 | 6.8e-2 | 2.9e-2 | 6.0e-2 | 1.7e-2 | 1.7e-2 |
| 480K | 3.3e+2 | 6.9e-1 | 2.4e-1 | 3.2e-1 | 8.2e-2 | 8.3e-2 |

Anticipating use in problems with photon counting processes, we assume Poisson models. The approach can be altered for Gaussian data models which may be a convenient approximation, employing the central limit theorem, when the number of counts for each detector is high. However, in scattering applications, the number of counts per detector is characteristically small (e.g. $\sim 100$) and the Poisson model is appropriate.

We have focused on a forward model (13) that has been used in many applications from transmission and emission tomography [40], to optical scatter imaging [1]. Our single-scatter approximation assumes the photon mean-free path is on the order of the system size [1]. This assumption, combined with collimation of the source and detector ensures most detected photons undergo a single scatter event. In addition, beam spreading and collimation will lead to spatial ambiguity in the scatter location. Our models do not account for spatial ambiguity explicitly, although our choice of regularization mitigates the de-blurring problem.

The choices of simulations in this paper were made to demonstrate the potential performance improvement in disparate cases. Scaling the attenuation image, source intensity, number of source locations, scatter angles, and transmission data all significantly affect the convergence rate and quality of the reconstructed images. Without transmission data, we observed ambiguity in the joint image estimation problem. To illustrate the ambiguity, we simulated images related quadratically using (36), (37). For low attenuating media, it is difficult to resolve the attenuation image due to its low signal strength in the data compounded by poor conditioning of the BRT operator. Resolving low attenuation images is particularly challenging from small scatter angles and low photon counts. Regions of high attenuation are also problematic as they can obscure other regions of interest leading to poor recovery and slow convergence. In application, the relationship between $\mu$ and $\alpha$ will be highly nonlinear which may further exacerbate detailed recovery of both images. The viability of joint image recovery should be investigated for known source attributes and media of interest. Our generalized algorithm provides a method to quantify performance from simulated data and the benefits of incorporating transmission measurements and multiple scatter angles for different media and sources.

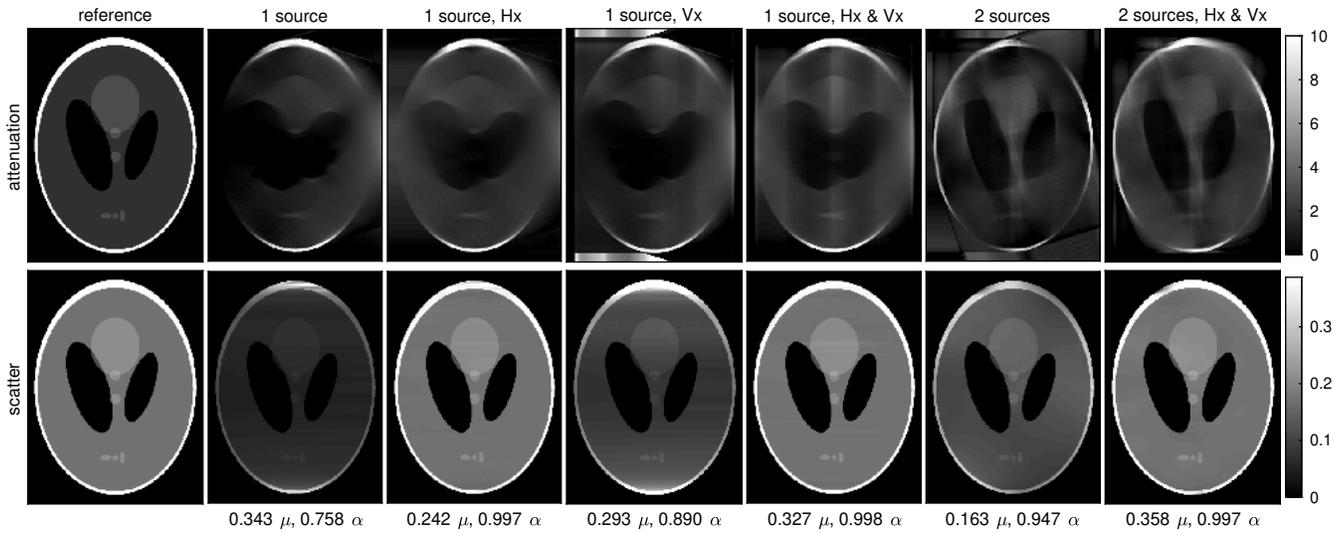

Figure 4. Contrasting measurement geometries with transmission measurements from noise-free data without regularization. Rows distinguish attenuation from scatter images. Each column contains an estimate from a different measurement geometry. Single source measurement geometries utilize two detectors, each detecting a symmetric scatter angle $\pm\pi/10$. For two-source measurement geometries, the sources are located at the directions $\pi$ and $\pi/2$, while the corresponding detectors are located at $\pi/10$ and $-\pi/2 + \pi/10$. We use Hx and Vx to represent horizontal and vertical transmission measurements associated with sources in the directions $\pi$ and $\pi/2$, respectively. The structural similarity index (SSIM) [39] is listed below each column for both images as indicated.

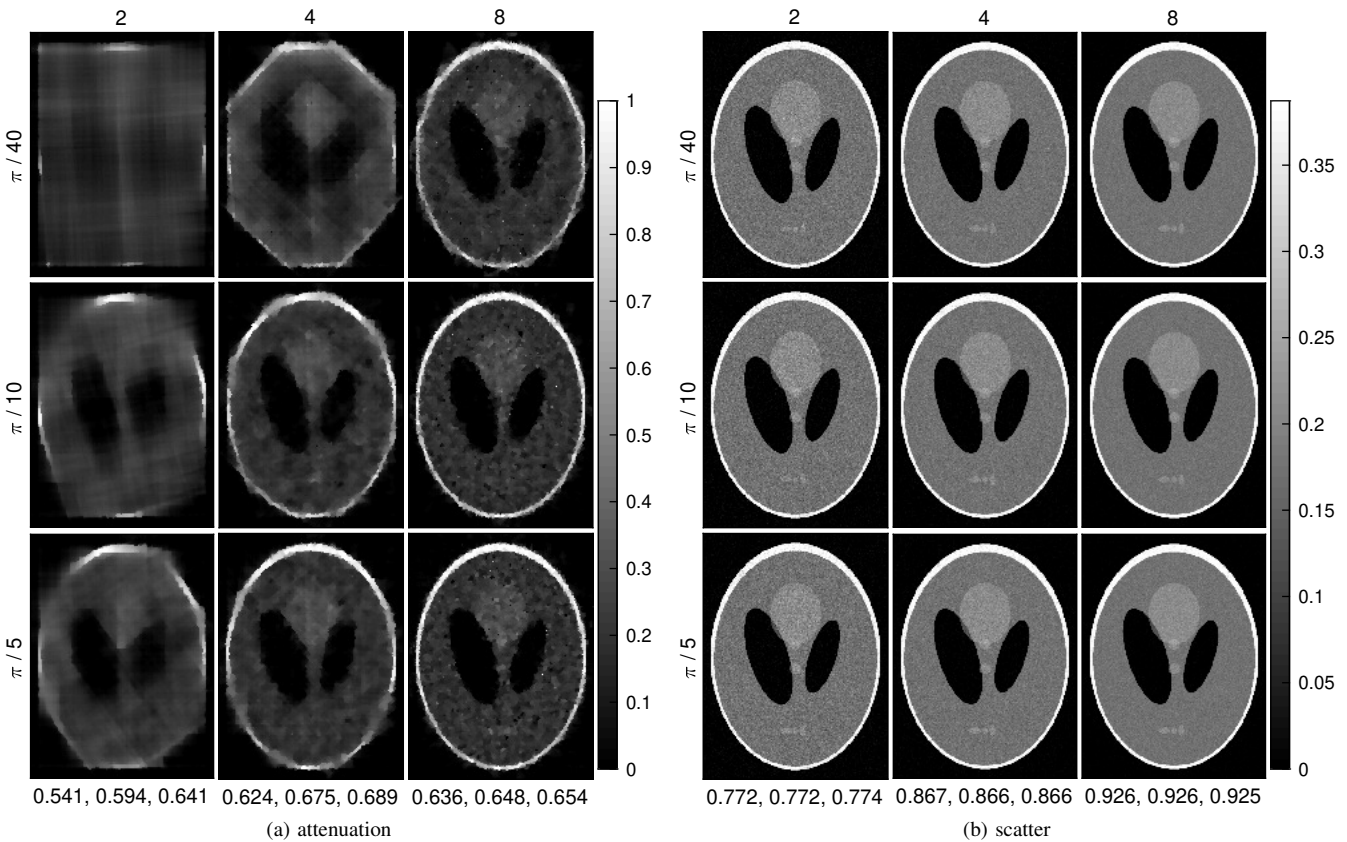

Figure 5. Effects of scatter angle and number of sources on image reconstruction. Attenuation and scatter images are shown in Fig. 5a and Fig. 5b, respectively. Rows distinguish scatter angles. Each column indicates a different number of sources. Each source direction is associated with a single detector at the indicated scatter angle. The source directions are uniformly distributed in angle between $\pm\pi/2$. Transmission measurements were used in both the horizontal and vertical directions. Bellow each column, the SSIM is listed for each image top to bottom.

Here we focus on the SVT measurement geometry, which we view as a necessary first step to addressing joint image reconstruction for coded aperture measurement geometries. The coded aperture data can be described as a sparse sampling across multiple SVT measurement geometries with differing scatter directions. Coded apertures simultaneously observe multiple scatter directions from each illuminated voxel. This is important for addressing the low signal counts associated with single-path single-scatter measurements. Previous work on coded apertures simplified the effects of the attenuation image as constant [12], or negligible [13], [14]. Our results suggest joint attenuation estimation, particularly with transmission data, could improve estimation of momentum transfer.

Joint image recovery for coded aperture measurement geometries will require computations in a high dimensional space comprising scattering from many angles. Ordered subsets is a traditional approach to decomposing iterative image reconstruction in high dimensions. Our simplification of the BRT forward and adjoint operator is particularly useful in this setting because it reduces the computational burden both constructing and applying the operator. However, our forward operator assumes global transform of a bounded image. Transforms of subsets must account for boundary conditions of the cone beam transform [3].

## APPENDIX A
## SCATTER IMAGE FIDELITY

Choosing $\mathcal{Y}$ to index both $\mathbf{d}$ and $\boldsymbol{\alpha}$ aides separability when updating scatter image estimates. Differentiating (16) with respect to $\alpha(y)$ we obtain separable functions for each $y \in \mathcal{Y}$. Further, (16) is convex with respect to $\boldsymbol{\alpha}$ for $\boldsymbol{\mu}$ fixed. Surrogate approximations are not necessary to update $\boldsymbol{\alpha}$. This is demonstrated through the following Lemma.

**Lemma 3.** *The I-divergence* (16) *is convex over* $\boldsymbol{\alpha} \in \mathcal{S}$ *when there exists at least one* $i \in \mathcal{I}$ *such that* $d_i(y) > 0$.

*Proof.* For convenience we define

$$\dot{g}_i(y:\boldsymbol{\mu}) \coloneqq \frac{\partial}{\partial \alpha(y)} g_i(y:\boldsymbol{\alpha},\boldsymbol{\mu}) \quad (41)$$

$$= I_0(y) \exp\left(-\sum_{x \in \mathcal{X}} h_i(y|x)\mu(x)\right) \quad (42)$$

We emphasize $\dot{g}_i(y:\boldsymbol{\mu})$ is independent of $\boldsymbol{\alpha}$. Since the BRT of the image is finite, and $I_0(y) > 0$, we have $\dot{g}_i(y:\boldsymbol{\mu}) > 0$ for all $y, i$.

Taking the derivative of (16) with respect to $\alpha(y)$, we find

$$\frac{\partial I(\mathbf{d} \parallel \mathbf{g}(\boldsymbol{\alpha},\boldsymbol{\mu}))}{\partial \alpha(y)} = \sum_{i \in \mathcal{I}} \left(\dot{g}_i(y) - \frac{d_i(y)\dot{g}_i(y:\boldsymbol{\mu})}{\alpha(y)\dot{g}_i(y:\boldsymbol{\mu}) + \beta_i(y)}\right). \quad (43)$$

The second derivative is then

$$\frac{\partial^2 I(\mathbf{d} \parallel \mathbf{g}(\boldsymbol{\alpha},\boldsymbol{\mu}))}{\partial \alpha(y)^2} = \sum_{i \in \mathcal{I}} \frac{d_i(y)\dot{g}_i(y:\boldsymbol{\mu})}{(\alpha(y)\dot{g}_i(y:\boldsymbol{\mu}) + \beta_i(y))^2}. \quad (44)$$

If $d_i(y)$ is positive for at least one $i \in \mathcal{I}$, then (44) is also positive. Therefore (16) is strictly convex over $\boldsymbol{\alpha} \in \mathcal{S}$. □

## APPENDIX B
## ATTENUATION FIDELITY SURROGATE

A surrogate for the data fidelity term (16) is available which is separable with respect to $x \in \mathcal{X}$. For this, we adopt the approach of O'Sullivan and Benac: recasting the problem as joint estimation over members of a linear family and an exponential family [40].

We first define two families of functions for expanding $\mathbf{d}$ and $\mathbf{g}$ as linear combinations. Let $\mathcal{L}(\mathbf{d})$ define a linearly family whose marginals equal the data

$$\mathcal{L}(\mathbf{d}) = \left\{\mathbf{p} : p_i(y,E) \geq 0, \quad \sum_E p_i(y,E) = d_i(y)\right\}. \quad (45)$$

Let $\mathcal{E}(\boldsymbol{\alpha})$ define an exponential family, associated with $\boldsymbol{\alpha}$, and parameterized by $\boldsymbol{\mu}$

$$\mathcal{E}(\boldsymbol{\alpha}) = \Big\{\mathbf{q} : q_i(y,0:\boldsymbol{\mu}) = \beta_i(y),$$
$$q_i(y,1:\boldsymbol{\mu}) = I_0(y)\alpha(y)\exp\left(-\sum_{x \in \mathcal{X}} h_i(y|x)\mu(x)\right)\Big\}. \quad (46)$$

The data model is composed by one member of the exponential family

$$g_i(y:\boldsymbol{\alpha},\boldsymbol{\mu}) = \sum_E q_i(y,E:\boldsymbol{\mu}). \quad (47)$$

In [40], $E$ was used to distinguish spectral measurements. However, the index remains useful for mono-energetic measurements when $\beta(y) > 0$.

The divergence between elements of the linear and exponential families reads

$$I(\mathbf{p} \parallel \mathbf{q}) = \sum_{i \in \mathcal{I}} \sum_{y \in \mathcal{Y}} \sum_E \left(p_i(y,E) \ln \frac{p_i(y,E)}{q_i(y,E:\boldsymbol{\mu})} - p_i(y,E) + q_i(y,E:\boldsymbol{\mu})\right). \quad (48)$$

Let $\hat{\mathbf{q}} \in \mathcal{E}(\boldsymbol{\alpha})$ indicate the element of the exponential family associated with $\hat{\boldsymbol{\mu}}$. Fixing $\hat{\mathbf{q}}$ in (48), we consider the minimizer $\mathbf{p} \in \mathcal{L}(\mathbf{d})$, which is subject to the linear constraint (45). The result is available in closed form

$$\hat{\mathbf{p}} = \underset{\mathbf{p} \in \mathcal{L}(\mathbf{d})}{\arg \min}\, I(\mathbf{p} \parallel \hat{\mathbf{q}}) \quad (49a)$$

$$= \left\{p_i(y,E) = d_i(y)\frac{q_i(y,E:\hat{\boldsymbol{\mu}})}{\sum_{E'} q_i(y,E':\hat{\boldsymbol{\mu}})}\right\}. \quad (49b)$$

Plugging this result back into (48), we find a variational form of (16),

$$I(\mathbf{d} \parallel \mathbf{g}(\boldsymbol{\alpha},\boldsymbol{\mu})) = \underset{\mathbf{p} \in \mathcal{L}(\mathbf{d})}{\min}\, I(\mathbf{p} \parallel \mathbf{q}), \quad (50)$$

due to O'Sullivan and Benac [40]. This motivates the surrogate function

$$D(\boldsymbol{\mu} : \hat{\boldsymbol{\mu}}) \coloneqq I(\hat{\mathbf{p}} \parallel \mathbf{q}), \quad (51)$$

where $\hat{\mathbf{p}}$, parameterized by $\hat{\boldsymbol{\mu}}$, is given by (49b). That (51) serves as a surrogate for (16) can be summarized as

$$D(\hat{\boldsymbol{\mu}} : \hat{\boldsymbol{\mu}}) = I(\mathbf{d} \parallel \mathbf{g}(\boldsymbol{\alpha},\hat{\boldsymbol{\mu}})) \quad (52)$$

$$D(\boldsymbol{\mu} : \hat{\boldsymbol{\mu}}) \geq I(\mathbf{d} \parallel \mathbf{g}(\boldsymbol{\alpha},\boldsymbol{\mu})), \quad \forall \boldsymbol{\mu} \in \mathcal{A}. \quad (53)$$

The equality in (52) is a restatement of (50). The inequality in (53) is an application of the convex decomposition lemma

$$f\left(\sum_x t(x)\right) \leq \sum_x r(x) f\left(\frac{t(x)}{r(x)}\right). \quad (54)$$

This holds for all $\mathbf{r} \in \{\mathbf{r} : r(x) \geq 0, \sum_x r(x) = 1\}$ and follows from Jensen's inequality [40].

Defining the auxiliary function

$$\psi_i(x|y) := h_i(y|x)(\mu(x) - \hat{\mu}(x)), \quad (55)$$

we use $\hat{\boldsymbol{\mu}}$ to parameterize $\hat{p}$ and $\hat{q}$ to restate

$$D(\boldsymbol{\mu} : \hat{\boldsymbol{\mu}}) = \sum_{i \in \mathcal{I}} \sum_{y \in \mathcal{Y}} \hat{p}_i(y,1) \sum_{x \in \mathcal{X}} (\psi_i(x|y) + h_i(y|x)\hat{\mu}(x))$$

$$+ \sum_{i \in \mathcal{I}} \sum_{y \in \mathcal{Y}} \hat{q}_i(y,1) \exp\left(-\sum_{x \in \mathcal{X}} \psi_i(x|y)\right) + d_0(\hat{\boldsymbol{\mu}}). \quad (56)$$

Here we have summarized all of the terms which do not depend on $\boldsymbol{\mu}$ with the single additive scalar $d_0(\hat{\boldsymbol{\mu}})$.

Recognizing the function

$$f(y, E, t) = t\hat{p}_i(y,1) + \hat{q}_i(y,1)\exp(-t) \quad (57)$$

as convex over $t$, we again make use of the convex decomposition lemma. This yields

$$D(\boldsymbol{\mu} : \hat{\boldsymbol{\mu}}) \leq d_0(\hat{\boldsymbol{\mu}}) + \sum_{i \in \mathcal{I}} \sum_{y \in \mathcal{Y}} \sum_{x \in \mathcal{X}} \left[\hat{p}_i(y,1)h_i(y|x)\mu(x)\right.$$

$$\left. + r_i(x|y)\hat{q}_i(y,1)\exp\left(-\frac{h_i(y|x)}{r_i(x|y)}(\mu(x)-\hat{\mu}(x))\right)\right] \quad (58)$$

for all $r_i(x|y) > 0$ such that

$$\sum_{x \in \mathcal{X}} r_i(x|y) = 1. \quad (59)$$

This constraint can be relaxed with the addition of a dummy $x = 0$ such that $h_i(y|0) = 0, \forall y, i$. This has no effect on (56), but $x = 0$ contributes to the right-hand side of (58). This bias is independent of $\boldsymbol{\mu}$ but varies with $\hat{\mathbf{q}}$.

We select

$$r_i(x|y) = \begin{cases} \dfrac{h_i(y|x)}{Z_i(x)}, & x \neq 0 \\ 1 - \displaystyle\sum_{x \in \mathcal{X} \setminus \{0\}} \dfrac{h_i(y|x)}{Z_i(x)}, & x = 0. \end{cases} \quad (60)$$

In general, $Z_i(x)$ must be sufficiently large such that $r_i(0|y) \geq 0$. This motivates the decoupled objective function

$$\overline{D}(\boldsymbol{\mu} : \hat{\boldsymbol{\mu}}) := \sum_{i \in \mathcal{I}} \sum_{y \in \mathcal{Y}} \sum_{x \in \mathcal{X} \setminus \{0\}} \left[\mu(x)\hat{p}_i(y,1)h_i(y|x)\right.$$

$$\left. + \hat{q}_i(y,1)\frac{h_i(y|x)}{Z_i(x)}\left(\exp\left(-Z_i(x)(\mu(x)-\hat{\mu}(x))\right) - 1\right)\right]$$

$$+ d_0(\hat{\boldsymbol{\mu}}) + \sum_{i \in \mathcal{I}} \sum_{y \in \mathcal{Y}} \hat{q}_i(y,1). \quad (61)$$

Here we incorporate additional terms due to $x = 0$ such that

$$\overline{D}(\hat{\boldsymbol{\mu}} : \hat{\boldsymbol{\mu}}) = D(\hat{\boldsymbol{\mu}} : \hat{\boldsymbol{\mu}}) \quad (62)$$

$$\overline{D}(\boldsymbol{\mu} : \hat{\boldsymbol{\mu}}) \geq D(\boldsymbol{\mu} : \hat{\boldsymbol{\mu}}), \quad \forall \boldsymbol{\mu} \in \mathcal{A}. \quad (63)$$

Combining these with (53) and (52), we find $\overline{D}$ is also a surrogate for $I(\mathbf{d} \| \mathbf{g}(\boldsymbol{\alpha}, \boldsymbol{\mu}))$.

The expression (61) can be simplified when $Z_i(x)$ is constant over $i$ and $x$. For this purpose, we define

$$Z_0 := \max_{y \in \mathcal{Y}, i \in \mathcal{I}} \sum_{x \in \mathcal{X}} h_i(y|x). \quad (64)$$

Further simplifying the notation, we define

$$b_0 := d_0(\hat{\boldsymbol{\mu}}) + \sum_{i \in \mathcal{I}} \sum_{y \in \mathcal{Y}} \hat{q}_i(y,1) \quad (65)$$

$$b_1(x) := \sum_{i \in \mathcal{I}} \sum_{y \in \mathcal{Y}} h_i(y|x)\hat{p}_i(y,1) \quad (66)$$

$$b_2(x) := \sum_{i \in \mathcal{I}} \sum_{y \in \mathcal{Y}} h_i(y|x)\hat{q}_i(y,1). \quad (67)$$

The expressions (66) and (67) comprise adjoint BRTs of $\hat{p}_i(y,1)$ and $\hat{q}_i(y,1)$, respectively. We restate (61)

$$\overline{D}(\boldsymbol{\mu} : \hat{\boldsymbol{\mu}}) = b_0 + \sum_{x \in \mathcal{X} \setminus \{0\}} \left[\mu(x)b_1(x)\right.$$

$$\left. + b_2(x)\frac{1}{Z_0}\left(\exp\left(-Z_0(\mu(x)-\hat{\mu}(x))\right) - 1\right)\right]. \quad (68)$$

The gradient separates as

$$\frac{\partial \overline{D}(\boldsymbol{\mu} : \hat{\boldsymbol{\mu}})}{\partial \mu(x)} = b_1(x) - b_2(x)\exp\left(-Z_0(\mu(x)-\hat{\mu}(x))\right). \quad (69)$$

The second derivative is nonnegative for all $\mu(x)$ since $b_2(x) \geq 0$. Therefore, $\overline{D}$ is convex with respect to $\boldsymbol{\mu}$.

## APPENDIX C
## REGULARIZATION SURROGATE

We generalize the regularization term for an image $\boldsymbol{\mu}$ and sample indices $\mathcal{X}$

$$R(\boldsymbol{\mu}) := \sum_{x \in \mathcal{X}} \sum_{z \in \mathcal{N}_x} w(x,z)\phi_\delta(\mu(x) - \mu(z)). \quad (70)$$

Here $\mathcal{N}_x \subset \mathcal{X}$ indicates the collection of voxels within a neighborhood of $x \in \mathcal{X}$, and $\phi_\delta : \mathbb{R} \to \mathbb{R}$ is an edge-preserving potential function. Specifically, we assume $\phi_\delta$ is strictly convex, even, and $\dot{\phi}_\delta(t)/t$ is monotone decreasing for $t > 0$. Therefore, $R(\boldsymbol{\mu}) \geq 0$ with equality for any constant image $\mu(x) = \mu_0$.

Using a constant image $\hat{\boldsymbol{\mu}}$, we expand $\phi_\delta(\mu(x) - \mu(z))$ using Jensen's inequality

$$\phi_\delta(\mu(x) - \mu(z)) \leq \frac{1}{2}\left[\phi_\delta(2\mu(x) - \hat{\mu}(x) - \hat{\mu}(z))\right.$$

$$\left. + \phi_\delta(2\mu(z) - \hat{\mu}(x) - \hat{\mu}(z))\right]. \quad (71)$$

Here we make use of both the convexity and symmetry of $\phi$. This motivates the separable surrogate regularization function

$$\check{R}(\boldsymbol{\mu} : \hat{\boldsymbol{\mu}}) := \frac{1}{2} \sum_{x \in \mathcal{X}} \sum_{z \in \mathcal{N}_x} w(x,z)$$

$$\times \left[\phi_\delta(2\mu(x) - \hat{\mu}(x) - \hat{\mu}(z)) + \phi_\delta(2\mu(z) - \hat{\mu}(x) - \hat{\mu}(z))\right] \quad (72)$$

due to De Pierro [41]. Separability is emphasized by restating

$$\breve{R}(\boldsymbol{\mu} : \hat{\boldsymbol{\mu}}) = \sum_{x \in \mathcal{X}} \breve{R}_x(\mu(x)), \quad (73)$$

where

$$\breve{R}_x(t) := \frac{1}{2} \sum_{z \in \mathcal{N}_x} w(x,z) \phi_\delta(2\mu(x) - \hat{\mu}(x) - \hat{\mu}(z)) \\ + \frac{1}{2} \sum_{z \in \mathcal{N}_x^b} w(z,x) \phi_\delta(2\mu(x) - \hat{\mu}(x) - \hat{\mu}(z)). \quad (74)$$

Here we use $\mathcal{N}_x^b := \{y : \mathcal{N}_y \ni x\}$ to represent the set of voxels with $x$ as a neighbor. For symmetric problems, when $\mathcal{N}_x = \mathcal{N}_x^b$ and $w(x,z) = w(z,x)$, the two terms in (74) are equivalent.

As a second approximation, we replace $\phi$ in (72) with the quadratic surrogate $\overline{\phi} : \mathbb{R} \to \mathbb{R}$,

$$\overline{\phi}(t) := \phi(\hat{t}) + \dot{\phi}(\hat{t})(t - \hat{t}) + \frac{1}{2} \frac{\dot{\phi}(\hat{t})}{\hat{t}} (t - \hat{t})^2 \quad (75)$$

$$\frac{\partial \overline{\phi}(t)}{\partial t} = \dot{\phi}(\hat{t}) + \frac{\dot{\phi}(\hat{t})}{\hat{t}}(t - \hat{t}). \quad (76)$$

This represents an upper bound on $\phi$ under the requirements $\phi$ is convex, symmetric, and when $\dot{\phi}(t)/t$ is monotone decreasing for $t > 0$. Equality is achieved at $t = \hat{t}$ such that $\overline{\phi}(\hat{t}) - \phi(\hat{t}) = 0$. Further, it can be shown $t = \hat{t}$ minimizes this difference (see Lemma 8.3 in [42]).

For the expansion point in (75), we use

$$\hat{t} = \hat{\mu}(x) - \hat{\mu}(z), \quad t - \hat{t} = 2(\mu(x) - \hat{\mu}(x)). \quad (77)$$

We define the separable quadratic surrogate

$$\overline{R}(\boldsymbol{\mu} : \hat{\boldsymbol{\mu}}) := \frac{1}{2} \sum_{x \in \mathcal{X}} \\ \left[ \sum_{z \in \mathcal{N}_x} w(x,z) \overline{\phi}_\delta(2\mu(x) - \hat{\mu}(x) - \hat{\mu}(z)) \\ + \sum_{z \in \mathcal{N}_x^b} w(z,x) \overline{\phi}_\delta(2\mu(x) - \hat{\mu}(x) - \hat{\mu}(z)) \right]. \quad (78)$$

Therefore, $\overline{R}(\boldsymbol{\mu} : \hat{\boldsymbol{\mu}}) \geq R(\boldsymbol{\mu})$ with equality when $\boldsymbol{\mu} = \hat{\boldsymbol{\mu}}$. For convenience, the following definitions are parameterized by $\hat{\boldsymbol{\mu}}$ and independent of $\boldsymbol{\mu}$:

$$c_0 := \frac{1}{2} \sum_{x \in \mathcal{X}} \left[ \sum_{z \in \mathcal{N}_x} w(x,z) \phi_\delta(\hat{\mu}(x) - \hat{\mu}(z)) \\ + \sum_{z \in \mathcal{N}_x^b} w(z,x) \phi_\delta(\hat{\mu}(x) - \hat{\mu}(z)) \right] \quad (79)$$

$$c_1(x) := \sum_{z \in \mathcal{N}_x} w(x,z) \dot{\phi}_\delta(\hat{\mu}(x) - \hat{\mu}(z)) \\ + \sum_{z \in \mathcal{N}_x^b} w(z,x) \dot{\phi}_\delta(\hat{\mu}(x) - \hat{\mu}(z)) \quad (80)$$

$$c_2(x) := \sum_{z \in \mathcal{N}_x} w(x,z) \frac{\dot{\phi}_\delta(\hat{\mu}(x) - \hat{\mu}(z))}{\hat{\mu}(x) - \hat{\mu}(z)} \\ + \sum_{z \in \mathcal{N}_x^b} w(z,x) \frac{\dot{\phi}_\delta(\hat{\mu}(x) - \hat{\mu}(z))}{\hat{\mu}(x) - \hat{\mu}(z)}. \quad (81)$$

Using these definitions in (78), we obtain

$$\overline{R}(\boldsymbol{\mu} : \hat{\boldsymbol{\mu}}) = c_0 + \sum_{x \in \mathcal{X}} \left[ c_1(x)(\mu(x) - \hat{\mu}(x)) \\ + c_2(x)(\mu(x) - \hat{\mu}(x))^2 \right]. \quad (82)$$

Taking the derivative with respect to $\mu(x)$, we find

$$\frac{\partial \overline{R}(\boldsymbol{\mu} : \hat{\boldsymbol{\mu}})}{\partial \mu(x)} = c_1(x) + 2c_2(x)(\mu(x) - \hat{\mu}(x)). \quad (83)$$

Observing $c_2(x) > 0$ for all $\hat{\boldsymbol{\mu}}$, $\overline{R}$ is strictly convex.